\documentclass{article}

\usepackage[preprint]{neurips_2026}

\usepackage[utf8]{inputenc} 
\usepackage[T1]{fontenc}    
\usepackage{hyperref}       
\usepackage{url}            
\usepackage{booktabs}       
\usepackage{amsfonts}       
\usepackage{nicefrac}       
\usepackage{microtype}      
\usepackage{xcolor}         
\usepackage{graphicx}
\usepackage{subcaption}
\usepackage{array}
\usepackage{tabularx}
\usepackage{booktabs}
\usepackage{amsmath} 
\title{DynoSys: A Dynamic Systems Framework for Multimodal Integration of Genetic, Environmental, and Neurobiological Signals}

\author{%
  Mengman Wei\thanks{Use footnote for providing further information
    about author (webpage, alternative address)---\emph{not} for acknowledging
    funding agencies.} \\
  Department of Neuroscience\\
  The Scripps Research Institute\\
  San Diego, CA 92108 \\
  \texttt{mwei@scripps.edu} \\
  \And
  Qian Peng \\
  Department of Neuroscience \\
  The Scripps Research Institute\\
  San Diego, CA 92108 \\
}

\begin{document}

\maketitle

\begin{abstract}
Understanding the development of adolescent behavioral and mental health outcomes requires integrating genetic predisposition, environmental exposures, and neurobiological processes over time. Here, we present a unified quantitative framework that models the human body as a dynamic system, where genetic factors form the foundational state, environmental exposures act as time-varying inputs, the brain might serve as a mediation processor, and behavioral phenotypes emerge as system outputs. Using longitudinal data from the Adolescent Brain Cognitive Development (ABCD) Study, we construct harmonized multi-domain representations across six phenotypes: externalizing behavior, internalizing behavior, and four substance-use initiation outcomes, including alcohol, nicotine, cannabis, and any substance use. We integrate polygenic risk scores, multi-domain environmental features, and multimodal neuroimaging representations derived through stability selection and dimensionality reduction. Our framework supports both continuous longitudinal modeling and survival-based event modeling through a unified data structure. We further develop interpretable domain-level representations using principal components, weighted risk scores, and cluster-based summaries. These representations enable downstream modeling using survival analysis, state-space models, and machine learning approaches. This work establishes a scalable and interpretable framework for studying how genetic and environmental factors interact over time to shape behavioral outcomes, providing a foundation for identifying modifiable risk factors and informing early intervention strategies.
\end{abstract}

\section{Introduction}

\begin{figure}[t]
    \centering
    \includegraphics[width=0.95\linewidth]{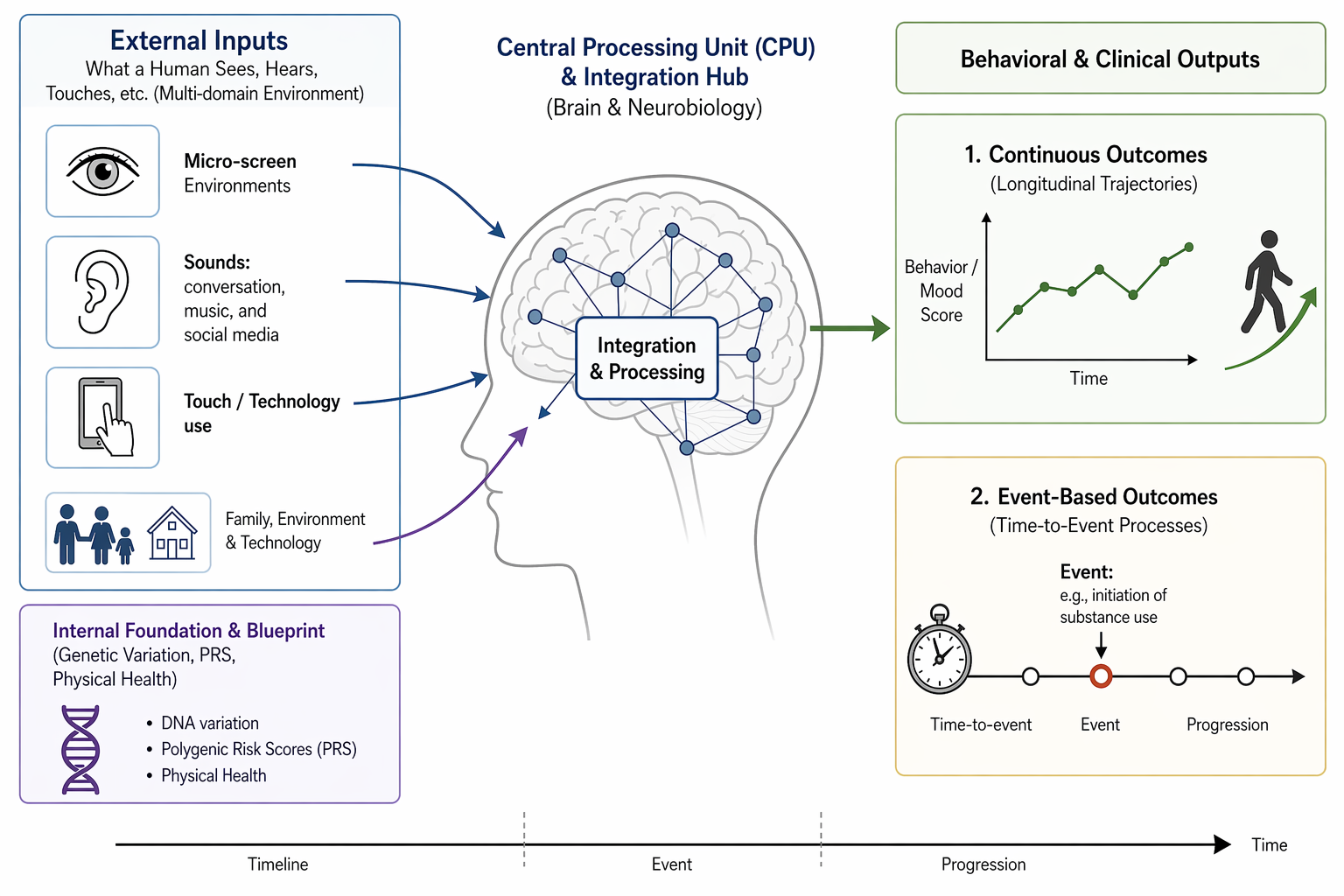}
    \caption{
    \textbf{Conceptual framework of human behavior as a dynamic system.}
    This figure shows human behavior as a system that changes over time and integrates multiple types of information. Genetic factors, such as DNA variation and polygenic risk scores (PRS) \cite{11,12,13,14}, provide a basic biological foundation. At the same time, external inputs, such as sensory experiences, family environment, social context, and technology use, continuously influence development. These internal and external inputs are processed by the brain, which acts as a central system that integrates information. The brain transforms these inputs into observable behaviors and clinical outcomes. We represent two types of outcomes: continuous trajectories, which describe gradual changes in behavior over time, and event-based outcomes, which capture specific transitions, such as the start of substance use, using time-to-event models. Overall, this framework highlights that behavior is dynamic and shaped by the interaction of genetic factors, environmental exposures, and brain processes over development.
    }
    \label{fig:conceptual_framework}
\end{figure}

Human behavior can be understood as the result of a complex dynamic system that evolves over time. This system is shaped by the interaction of biological, environmental, and experiential factors across development. Genetic variation provides a foundational biological basis, influencing brain structure and function as well as individual differences in susceptibility to behavioral outcomes. At the same time, environmental exposures, such as family context, social interactions, and broader physical and social environments, continuously influence and modify these biological processes. The brain plays a central role in this system. It integrates information from both internal biological signals and external environmental inputs, and it generates behavioral and clinical outcomes. In this sense, behavior is not determined by a single factor, but emerges from the ongoing interaction between multiple domains, as illustrated in Figure~\ref{fig:conceptual_framework}.

This systems-based view is consistent with established frameworks in systems neuroscience, developmental psychopathology, and the bioecological model of human development, which emphasize that behavior arises from interactions across multiple levels rather than from isolated factors \cite{1,2,3,6,7}. Similar concepts have been explored in prior work integrating genetics, environment, and brain imaging to study complex traits, although many existing approaches still analyze these components separately \cite{4,8,9}. Existing frameworks conceptualize human behavior as the result of multi-level interactions across biological and environmental systems.

Recent advances in imaging genetics and large-scale neuroimaging studies have enabled the integration of genetic and neurobiological data \cite{4,9}. In parallel, developments in machine learning have made it possible to model high-dimensional, multimodal data across domains \cite{10}. However, many of these approaches remain limited by cross-sectional designs or focus on a single outcome type. In contrast, our framework provides a unified, time-aware system that integrates genetic, environmental, and neurobiological factors to jointly model both longitudinal trajectories and event-based outcomes.

Despite major progress in large-scale genome-wide association studies (GWAS) and environmental epidemiology, most existing methods treat behavioral outcomes as static variables. These outcomes are often analyzed as cross-sectional or binary traits, which do not reflect the longitudinal and time-dependent nature of human behavior, particularly in psychiatric conditions and substance use \cite{5,9}. In addition, genetic, environmental, and neurobiological factors are typically studied separately. This limits our ability to understand how these systems jointly influence behavior across development.

To address these challenges, we propose a unified framework based on three main ideas.

\paragraph{Human behavior as a dynamic system.}
We view the human body as a system where behavior is the final output. In this system, inputs include different types of environmental exposures. Genetic factors, such as polygenic risk scores (PRS) \cite{11,12,13,14,15}, provide a foundation for biological states, including physical health and brain function. The brain and biological systems act as processors that integrate these inputs. The outputs are observable behaviors and clinical outcomes.

\paragraph{Modeling time explicitly.}
Time is a key part of our framework. We include time in two ways. First, longitudinal trajectories are used to describe how continuous behavioral variables change across development. Second, time-to-event models are used to capture key transitions, such as substance use initiation, for binary event-based variables in the available dataset.

\paragraph{Integration across domains.}
We combine multiple types of data to reflect the complexity of real-world risk. These include genetic data, represented by polygenic risk scores from GWAS; environmental data, including family, socioeconomic, physical, and technology-related factors; and neurobiological data, including brain imaging features that capture brain structure and function.

We apply this framework to six phenotypes using data from the Adolescent Brain Cognitive Development (ABCD) Study \cite{16}, a large longitudinal cohort that follows 8- to 9-year-old children across development with detailed behavioral, environmental, and neuroimaging data. The phenotypes include externalizing behavior, internalizing behavior, alcohol use initiation, nicotine use initiation, cannabis use initiation, and any substance use initiation.

This approach allows us to model both continuous behavioral changes over time and event-based outcomes within a single pipeline. Our goal is to understand how genetic and environmental factors work together to influence brain development and behavior across development.

\begin{figure}[t]
    \centering
    \includegraphics[width=0.95\linewidth]{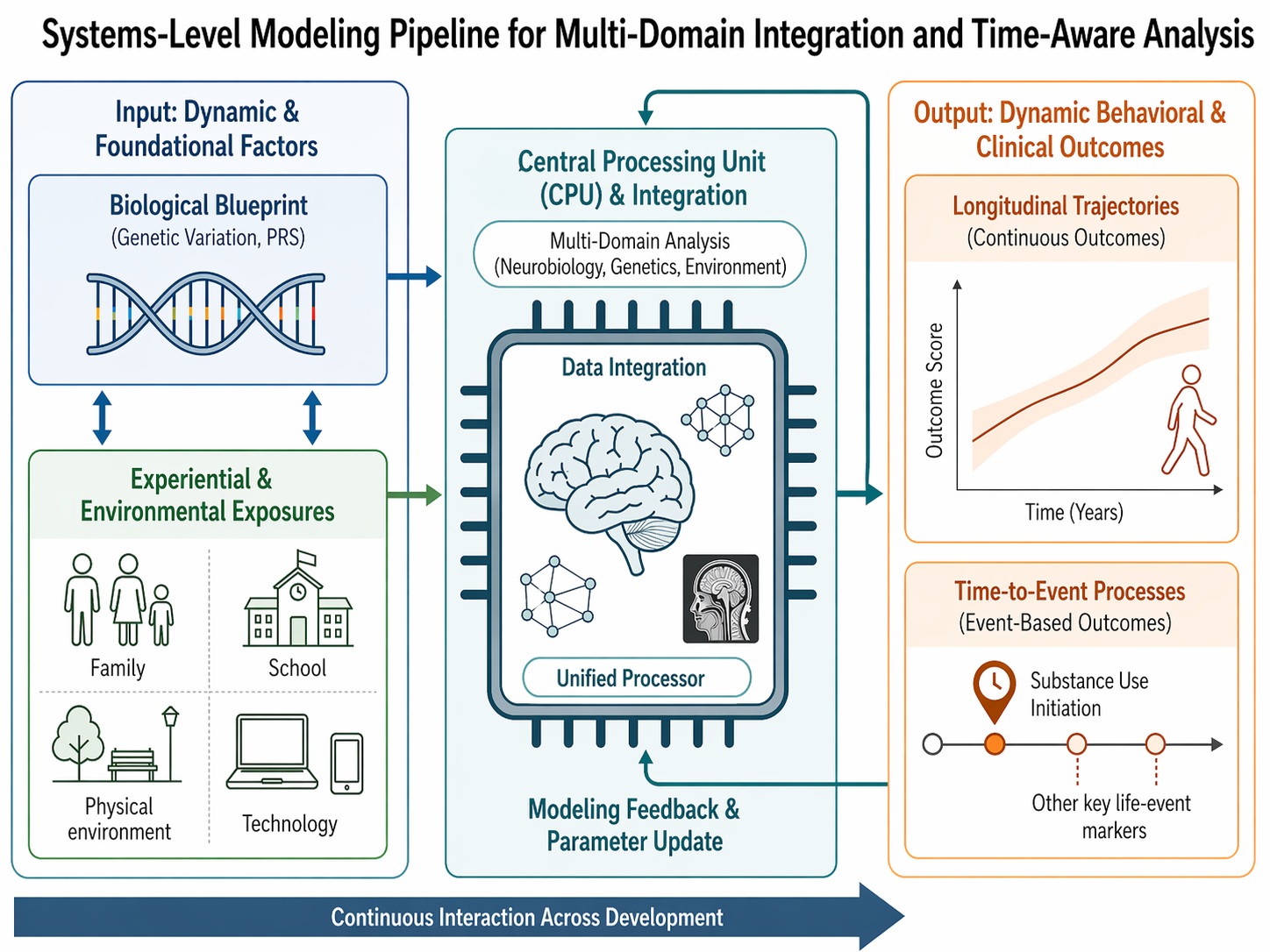}
    \caption{
    \textbf{Systems-level modeling pipeline for multi-domain integration and time-aware analysis.}
    Overview of the proposed systems-level modeling pipeline for understanding human behavior through multi-domain integration and time-aware analysis. The framework begins with two primary input domains: biological foundations, including genetic variation and polygenic risk scores (PRS), and dynamic environmental and experiential exposures, including family, school, physical, and technology-related contexts. These heterogeneous inputs are integrated within a central processing module that performs unified, multimodal analysis across genetic, environmental, and neurobiological data, such as brain imaging features. The integrated representations are then used to model two classes of outcomes: longitudinal trajectories of continuous behavioral phenotypes, capturing developmental changes over time, and time-to-event processes, modeling discrete behavioral transitions such as substance use initiation and other key life events. The framework incorporates iterative feedback and parameter updating to refine model performance and improve representation learning. This unified pipeline is applied to multiple phenotypes in the ABCD Study, enabling consistent and scalable modeling of both continuous and event-based behavioral outcomes within a single analytical framework.
    }
    \label{fig:systems_pipeline}
\end{figure}

\paragraph{General interpretation of the framework.}
This framework can be summarized as a simple system model, as shown in Figure~\ref{fig:systems_pipeline}. Genes provide the biological foundation. Environmental inputs contribute additional information to the system. Together, these inputs form the basis of a generative model. The brain acts as an inference system that processes and integrates information. Behavior is the output of this process.

We model how this system changes over time. Longitudinal modeling captures how the system evolves across development, while time-to-event modeling captures key transitions, such as substance use initiation, under uncertainty.

\section{Methods}

\subsection{Study Design Overview}

We developed a multi-stage pipeline to construct a unified longitudinal dataset integrating multiple domains, including genetic data, environmental factors, neuroimaging measures, and behavioral outcomes. Genetic information was represented using polygenic risk scores (PRS), environmental factors were represented using multi-domain exposure features, neuroimaging measures were represented using brain-derived features, and behavioral outcomes included both continuous and time-to-event outcomes. All data were organized into a time-indexed panel structure for each individual, allowing consistent tracking of changes over time (Figure~\ref{fig:systems_pipeline}).

The pipeline was designed as a systems-level framework for integrating multi-domain data and performing time-aware analysis. The framework starts with two main input domains: biological factors, including genetic variation and PRS, and environmental and experiential exposures, including family, school, physical environment, and technology use.

These inputs were combined in a central analysis module that integrates genetic, environmental, and neurobiological data into unified representations. These representations were then used to model two types of outcomes. First, continuous behavioral trajectories were modeled to capture changes in behavior over time. Second, time-to-event models were used to study discrete outcomes, such as substance use initiation and other key events.

The framework also included iterative updating, where model parameters and representations were refined as more information was incorporated. We applied this unified pipeline to multiple phenotypes in the ABCD Study, enabling consistent and scalable analysis of both continuous and event-based behavioral outcomes within a single framework.

\subsection{Outcome Definitions}

\subsubsection{Continuous Outcomes}

We defined two continuous behavioral outcomes: externalizing behavior and internalizing behavior. Both outcomes were measured using T-scores derived from the Child Behavior Checklist (CBCL) assessment in the ABCD Study.

These outcomes were organized in a longitudinal format. Each individual had repeated measurements across multiple time points, including baseline, 1-year follow-up, 2-year follow-up, and later waves when available. To capture temporal dynamics, we included lagged outcomes, representing the value of the outcome at the previous time point, and change features, representing the difference between consecutive time points. All continuous outcomes were constructed and processed within a unified pipeline.

\subsubsection{Survival Outcomes}

We defined four event-based outcomes: alcohol initiation, nicotine initiation, cannabis initiation, and any substance use initiation, following a recent substance use initiation study \cite{17}. These outcomes were modeled as time-to-event processes. For each individual, we tracked the time from baseline age to the first occurrence of the event.

The modeling used baseline age combined with follow-up time to define the time scale, and interval-based event indicators to represent whether the event occurred within each time interval. We constructed these outcomes using an anchor-based approach, with fixed time points at 0, 24, and 48 months. This ensured consistent alignment of follow-up intervals across individuals. For each outcome, case status was defined based on whether initiation occurred within the observed follow-up period, while individuals without events were treated as censored \cite{17}.

\subsection{Ancestry Definition and Analysis Sample}

PRS analyses were restricted to individuals of European ancestry to match the ancestry of the discovery GWAS. European ancestry was defined using a genotype-based approach rather than self-reported race. Specifically, ABCD samples were projected onto the 1000 Genomes reference panel \cite{18}, and ancestry was assigned using principal component analysis (PCA) followed by k-nearest-neighbor classification in genetic principal component space. Individuals with a high proportion of European neighbors were classified as European. From this group, unrelated individuals were further identified using a curated unrelated sample list provided by ABCD.

\subsection{PRS Construction}

We constructed PRS for three domains. Externalizing PRS were based on large-scale multivariate GWAS of externalizing behaviors \cite{14,15}. Internalizing PRS included PRS for major depressive disorder \cite{19}, anxiety disorders \cite{20}, neuroticism \cite{21}, and youth internalizing symptoms \cite{22}. To improve robustness in this youth sample, we additionally constructed a composite internalizing PRS by integrating multiple related traits. Substance use PRS included PRS for problematic alcohol use \cite{23}, cannabis use disorder \cite{24}, nicotine and alcohol use \cite{25}, and general substance use disorder liability \cite{26}.

For each GWAS, summary statistics were harmonized and reformatted for PRS-CS \cite{27}. Posterior SNP effect sizes were estimated separately for each chromosome using LD information from an external reference panel and then combined into a genome-wide score. PRS were calculated using PLINK 1.9 \cite{28} by summing allele dosages weighted by PRS-CS effect sizes. All PRS were computed at the individual level, aligned to the same genotype reference panel, and standardized prior to downstream analyses when appropriate.

\subsection{Multi-Domain Feature Representation Modeling}

Environmental and behavioral features were obtained from the ABCD dataset and organized into predefined domains, including environment, family context, physical health, technology use, and neurocognition. These domains were defined to capture different aspects of a child's developmental context, including external exposures and internal functional characteristics. Detailed descriptions of variables and domain assignments are provided in the Supplementary Tables. The same feature construction and selection procedure was applied to all domains to ensure consistency across analyses \cite{29}.

To build reliable and interpretable features, we developed a multi-stage pipeline for longitudinal data. The goal was to reduce noise, handle missing data, and identify stable signals related to the outcomes. The pipeline included three main steps: data screening and quality control, statistical feature selection, and representation learning.

In Stage 1, all variables were evaluated for data quality. Variables with high levels of missingness were removed based on predefined thresholds. For longitudinal variables, we assessed how often they were observed across follow-up waves and required a minimum level of activity over time. Based on these patterns, variables were classified as either longitudinal or static-like. For variables with limited variation across time but acceptable data coverage, missing values were filled using last observation carried forward (LOCF) within each individual \cite{30}. We also created lagged variables and change variables to capture within-individual temporal patterns. These steps ensured that only variables with sufficient data quality and longitudinal relevance were included in later analyses.

In Stage 2, we tested the association between each candidate variable and the outcome. Linear regression models were used for continuous outcomes, and logistic regression models were used for binary outcomes. All models included basic covariates such as time and, when appropriate, prior outcome values. To improve stability, we applied cross-validation by splitting the data at the individual level into multiple folds \cite{31}. For each variable, we evaluated how often it showed a significant association across folds and whether the direction of effect was consistent. Variables were selected if they met predefined criteria for selection frequency and sign consistency. This reduced overfitting and ensured that selected features were stable and reproducible.

In Stage 3, selected variables were summarized into domain-level representations. PCA was applied to capture the main patterns of variation and reduce dimensionality \cite{32}. In addition, we constructed aggregated scores by combining variables using estimated effect sizes, as well as simple averages. Variables were also grouped into risk and protective sets based on the direction of their associations with the outcome. From these groups, we calculated summary scores to represent overall risk and protective effects within each domain. These representations provide interpretable and compact summaries of complex features.

The final output of this pipeline consisted of domain-level features that combine longitudinal structure, statistical relevance, and reduced dimensionality. These features were used as inputs for downstream models, allowing unified and scalable analysis of multiple domains.

\subsection{Neuroimaging Feature Processing and Quality Control}

Structural MRI data from the ABCD Study were processed using a standardized imaging pipeline \cite{29,33}. We included three time points: baseline, 2-year follow-up, and 4-year follow-up. Imaging features were extracted across multiple structural modalities, including cortical area, sulcal depth, cortical thickness, and volumetric measures. These features were derived using well-established surface-based and volumetric processing methods implemented in FreeSurfer \cite{34}.

All imaging tables were first filtered to retain only the predefined time points. Features from different modalities were then merged by subject and visit to form a unified longitudinal dataset. Intracranial volume was included and used to normalize volumetric features to account for head size differences.

Quality control procedures were applied at multiple levels following established practices for large-scale neuroimaging studies \cite{35}. We excluded scans that failed inclusion flags for T1-weighted images, had low image quality scores, or had invalid intracranial volume values. Additional QC metrics related to topological defects, rigid registration, and nonlinear warping were evaluated, and extreme values were identified using high quantile thresholds. Two QC definitions were used: a primary QC based on basic inclusion criteria, and a stricter QC including all QC filters. This allowed flexibility in downstream analyses.

To ensure data reliability, features with high missingness greater than 20\% were removed. The remaining features were grouped by modality, including area, sulcal depth, thickness, and intracranial-volume-normalized volume, and a feature manifest was saved for reproducibility. The final output was a longitudinal dataset containing cleaned imaging features, QC indicators, scanner-related variables, and covariates such as sex, age, site, and genetic principal components, which are commonly used to control for population structure \cite{36}.

\subsection{Construction of Time-Aware Imaging Features}

To capture temporal dynamics, we constructed time-aware imaging features using a landmark-based design, which is commonly used in longitudinal and survival analyses \cite{37}. Three analysis settings were defined: baseline, 24-month landmark, and 48-month landmark.

For the baseline analysis, only imaging features at baseline were used. For the 24-month and 48-month landmark analyses, we included only individuals who had not yet experienced the event before the landmark time. Imaging features from multiple time points were combined to capture both static and dynamic information.

Specifically, we constructed baseline features, follow-up features, change features representing differences between time points, and slope features representing rates of change over time. These derived features allowed the model to capture both the current brain state and its developmental trajectory, which is important in longitudinal neuroimaging studies \cite{38}. All features were combined with baseline covariates, including sex, age, site, and principal components, and aligned with time-to-event outcomes to form the final design matrix.

\subsection{Stability-Based Feature Selection Using Cox LASSO}

We applied a stability selection framework based on Cox proportional hazards models \cite{38} with LASSO regularization \cite{39,40} to identify robust brain imaging features associated with time-to-event outcomes.

Before model fitting, imaging features were residualized with respect to nuisance covariates, including demographic variables and genetic principal components. This step removed confounding effects and ensured that selected features reflected associations beyond basic covariates \cite{36}.

Feature preprocessing included removal of features with high missingness, median imputation for missing values, variance filtering to remove near-constant features, standardization, and correlation pruning to remove highly correlated features.

We then performed repeated subsampling stability selection \cite{41}. In each iteration, a random subset of the data was used to fit a Cox LASSO model, and the optimal regularization parameter was selected based on validation concordance index. Features selected across iterations were tracked, and stability scores were defined as the proportion of times a feature was selected. Features with stability above a predefined threshold were retained as stable predictors. This approach reduced overfitting and ensured that selected features were reproducible across subsamples.

\subsection{Brain Representation Learning}

To obtain compact and interpretable summaries of brain imaging features, we developed a representation learning framework based on stable features identified in the previous step.

Stable features were first grouped by imaging modality and time block. For each group, we constructed three types of representations. First, PCA was applied to standardized features to capture the main sources of variation and reduce dimensionality \cite{32}. The first few principal components were used as low-dimensional summaries. Second, weighted scores were constructed based on survival associations. For each feature, we estimated its association with the outcome using univariate Cox models, and weights were defined based on estimated coefficients and statistical significance. Third, features were divided into risk and protective groups based on positive or negative associations with the outcome. For each group, we computed average scores, as well as a net score defined as the difference between risk and protective components.

These representations provide complementary views of brain structure, capturing overall variation, outcome-relevant signals, and directional effects.

\subsection{Adaptive Feature Compression and Integration}

To balance interpretability and model complexity, we applied an adaptive compression rule when constructing final brain inputs. When the number of stable features was small, both raw features and derived representations were retained. When the number of stable features was large, only compact representations were used.

The final brain representations were organized into a standardized format and merged with other domains, including genetic and environmental features. Each representation was indexed by individual and time, ensuring compatibility with the longitudinal panel structure used throughout the framework. These brain-derived features served as the neurobiological component of the system, linking genetic predisposition and environmental exposures to behavioral outcomes.

\subsection{Unified Data Integration}

To support multimodal modeling across genetic, environmental, neurobiological, and behavioral domains, we constructed a unified longitudinal dataset that integrated all available data sources into a single analysis-ready table. This process ensured consistent subject alignment, time indexing, and feature representation across heterogeneous inputs.

We first defined a common longitudinal anchor panel for each phenotype. For continuous outcomes, such as externalizing and internalizing scores, we used repeated measurements across study waves and mapped each observation to a discrete time index corresponding to baseline and follow-up visits. For binary outcomes, such as substance initiation, we constructed anchor time points at baseline, 24 months, and 48 months, and determined whether an event had occurred by each anchor. This approach ensured a consistent temporal structure across both continuous and time-to-event settings.

All datasets were harmonized using a shared set of identifiers, including individual ID and event name, and aligned to the same time index. Only valid time points were retained, and all records were sorted by subject and time to preserve temporal order. For continuous outcomes, we additionally created lagged variables and change scores to capture temporal dynamics.

Genetic information was incorporated using PRS. For each phenotype, PRS values were merged with the anchor panel using individual IDs. In cases where multiple related PRS measures were available, such as internalizing traits, we standardized each score and constructed a combined PRS by averaging across standardized components. This provided a stable representation of genetic liability while preserving individual-level variation.

We then integrated domain-specific predictors from multiple sources, including environmental factors, family context, physical health, technology use, and neurocognitive measures. Each domain had been previously processed into standardized feature tables. These tables were merged with the anchor panel using individual ID, event name, and time index. To avoid duplication and ensure consistency, overlapping variables were removed before merging. All domain features were retained in their processed form, allowing flexible comparison across different representation strategies.

Neuroimaging features were added in a separate step. Imaging data from baseline and follow-up time points were aligned to the corresponding anchor events. Multiple representations of brain features were supported, including raw features and derived representations such as principal components, weighted summaries, and clustered features. These representations were merged sequentially while ensuring that no duplicate columns were introduced.

Demographic and technical covariates, including sex, age, principal components, and site indicators, were also incorporated. Site effects were encoded using one-hot variables to allow adjustment during modeling. All covariates were converted to numeric format to ensure compatibility with downstream machine learning models.

After merging all components, the final dataset was cleaned and standardized. Duplicate columns were removed, and all non-identifier variables were converted to numeric values. The dataset was then sorted by subject and time index to maintain a consistent longitudinal structure.

To support systematic evaluation, we generated multiple versions of the unified dataset corresponding to different feature representation strategies, including PCA-based, weighted, clustered, and full feature sets. For each version, we also created a manifest file that recorded the domain origin of each variable. This enabled transparent tracking of features and facilitated interpretation of model results.

The resulting unified tables provided a consistent and flexible foundation for downstream modeling. They allowed simultaneous analysis of genetic, environmental, brain, and behavioral data within a single framework, while preserving temporal structure and subject-level consistency.

\subsection{Modeling Framework}

\subsubsection{Overview}

The unified dataset was used to train a series of models with increasing complexity. All models followed a consistent temporal design, where predictors from previous time points were used to predict future outcomes. Specifically, lagged variables were constructed so that predictors at time $t-1$ were used to predict outcomes at time $t$. This ensured correct temporal ordering and avoided information leakage.

For each phenotype, multiple versions of the dataset were constructed using different feature representations, including PCA, weighted scores, risk/protective clusters, and the full feature set. All models were trained separately on each representation to evaluate robustness and consistency.

\subsubsection{Data Representation and Time Structure}

All models operated on longitudinal panel data. For each individual $i$ at time $t$, the dataset included predictors $X_{i,t}$ and outcomes $Y_{i,t}$. To enforce temporal ordering, predictors were shifted by one time step:
\begin{equation}
X_{i,t-1} \rightarrow Y_{i,t}.
\end{equation}

For forward prediction tasks, we also defined:
\begin{equation}
Y_{i,t+1} = \text{outcome at the next time point},
\end{equation}
and
\begin{equation}
\Delta Y_{i,t+1} = Y_{i,t+1} - Y_{i,t}.
\end{equation}

Binary event indicators were defined as:
\begin{equation}
E_{i,t} =
\begin{cases}
1, & \text{if } Y_{i,t+1} \text{ exceeds a predefined threshold},\\
0, & \text{otherwise}.
\end{cases}
\end{equation}

These definitions allowed both continuous prediction and event-based modeling.

\subsection{Classical Models}

\subsubsection{Linear Models for Continuous Outcomes}

We used penalized linear models, including ridge regression, LASSO, and elastic net, to predict future outcomes:
\begin{equation}
Y_{i,t+1} = \beta_0 + \boldsymbol{\beta}^{\top} X_{i,t} + \epsilon_{i,t}.
\end{equation}

To improve robustness, covariates were separated from main predictors, non-covariate features were residualized with respect to covariates, and grouped cross-validation was performed at the subject level. This ensured that repeated measurements from the same individual did not bias the model.

\subsubsection{Discrete-Time Hazard Models}

For event prediction, we used a discrete-time survival formulation implemented as logistic regression:
\begin{equation}
P(Y_{i,t}=1 \mid X_{i,t}) =
\frac{1}{1 + \exp[-(\beta_0 + \boldsymbol{\beta}^{\top} X_{i,t})]},
\end{equation}
where $Y_{i,t}$ indicates whether an event occurred in the next interval. This formulation allowed modeling of substance initiation and behavioral worsening events. Model performance was evaluated using AUROC, AUPRC, and Brier score.

Although implemented as discrete-time models, this formulation is related to the Cox proportional hazards model:
\begin{equation}
h(t \mid X) = h_0(t)\exp(\boldsymbol{\beta}^{\top}X),
\end{equation}
providing a connection between classical survival analysis and machine learning models.

\subsection{Advanced Models}

\subsubsection{Multimodal Neural Survival Model}

We implemented a multimodal neural network to integrate different data domains. Each modality was processed through a separate branch:
\begin{equation}
h_m = f_m(X_m),
\end{equation}
where $X_m$ represents predictors from modality $m$ and $f_m$ is a modality-specific neural network. All modality-specific representations were then combined:
\begin{equation}
h_{\mathrm{fusion}} = \mathrm{concat}(h_1, h_2, \ldots, h_M).
\end{equation}
The final prediction was:
\begin{equation}
\hat{Y} = g(h_{\mathrm{fusion}}),
\end{equation}
where $g$ is the final prediction layer.

Modalities included genetic PRS, environment, family context, physical health, technology use, neurocognition, and brain imaging. This structure allowed the model to learn interactions across domains.

\subsubsection{State-Space Model}

To capture dynamic processes, we constructed a latent state-space model. State evolution was defined as:
\begin{equation}
z_t = A z_{t-1} + B x_{t-1} + \epsilon_t,
\end{equation}
and the observation model was defined as:
\begin{equation}
Y_t = C z_t + D x_{t-1} + \eta_t,
\end{equation}
where $z_t$ is the latent internal state, $x_t$ represents observed inputs, and $Y_t$ is the observed outcome. Latent states were derived using PCA on internal features such as brain and neurocognition. Bootstrapping was used to estimate stable transition and outcome effects.

\subsubsection{SEM-Style Causal Interpretation}

We implemented a longitudinal mediation framework. For predictor $X$, mediator $M$, and outcome $Y$, the mediation model was written as:
\begin{equation}
M = aX + \epsilon_M,
\end{equation}
\begin{equation}
Y = c'X + bM + \epsilon_Y.
\end{equation}
The indirect effect was defined as:
\begin{equation}
\mathrm{ACME} = a \times b.
\end{equation}
Bootstrapping was used to estimate confidence intervals for mediation effects. This provided interpretable pathways linking genetic factors, environmental exposures, internal states, and behavioral outcomes.

\subsection{Important Design Clarifications}

All models used all available longitudinal time points. The data were structured as panel data, and models learned transitions over time rather than relying on a single time point such as baseline or 24-month follow-up.

All feature representations were used, including PCA representation, weighted representation, cluster-based representation, and the full feature set. Models were trained on each representation separately.

Brain imaging features were treated as time-varying predictors when available. They were aligned to the longitudinal structure and used as lagged predictors. Overall, all models were trained across multiple feature representations and all available longitudinal time points, allowing consistent evaluation of predictive performance and temporal dynamics.

\subsection{Multi-Task Learning Extension}

\begin{figure}[t]
    \centering
    \includegraphics[width=\linewidth]{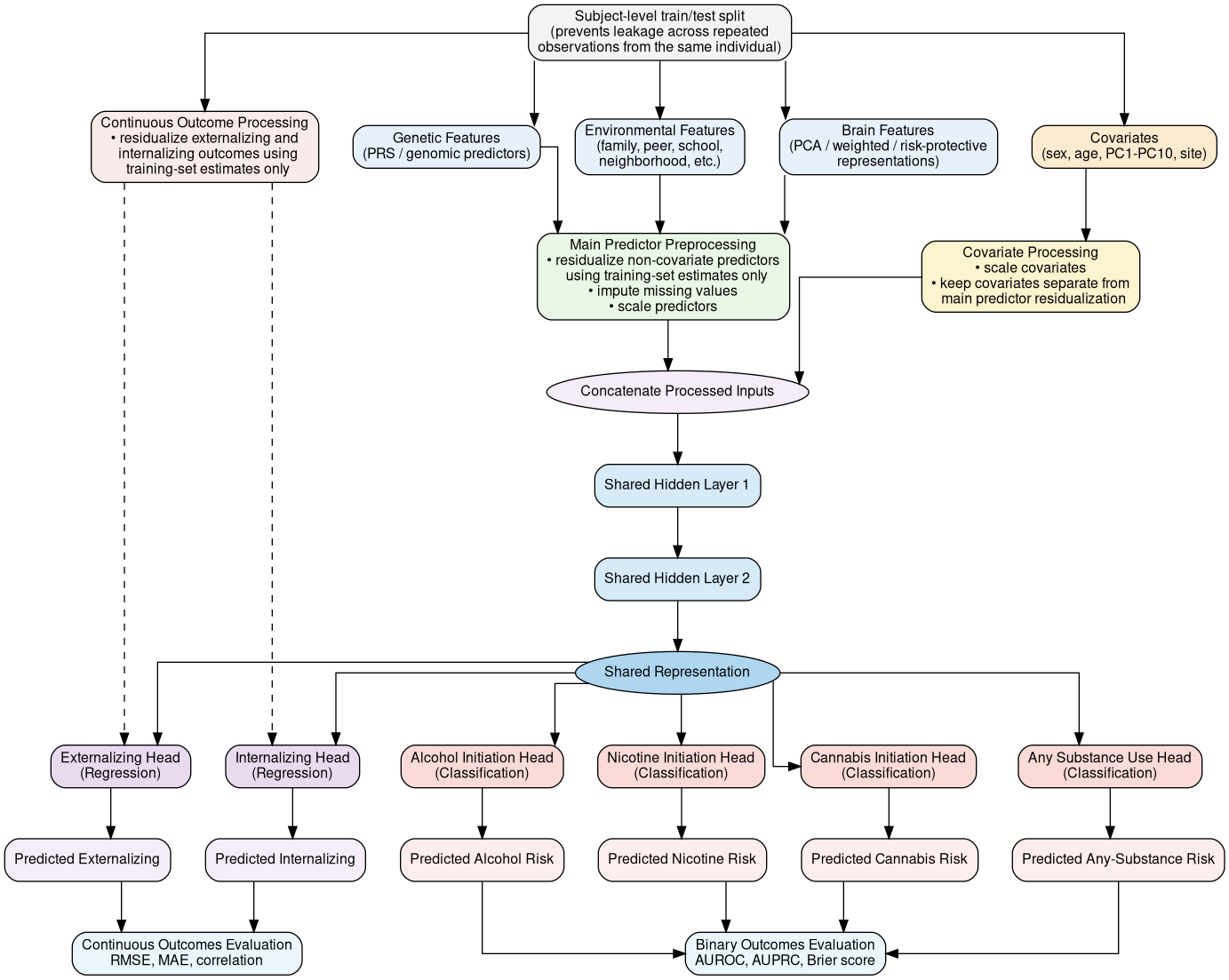}
    \caption{\textbf{Multi-task learning architecture.} Multimodal predictors, including genetic, environmental, and brain-derived features, were preprocessed separately from covariates. Non-covariate predictors were residualized against covariates using training-set estimates only, then imputed and scaled. Covariates were processed separately and concatenated with residualized predictors before entering a shared-bottom neural network. The network learned a common latent representation and then branched into six task-specific heads: two regression heads for externalizing and internalizing outcomes, and four classification heads for alcohol, nicotine, cannabis, and any-substance initiation. Continuous outcomes were evaluated using RMSE, MAE, and correlation, while binary outcomes were evaluated using AUROC, AUPRC, and Brier score.}
    \label{fig:multitask}
\end{figure}

We extended the modeling framework to a multi-task learning setting (Figure~\ref{fig:multitask}) so that shared and phenotype-specific patterns could be learned jointly across six outcomes: externalizing behavior, internalizing behavior, alcohol initiation, nicotine initiation, cannabis initiation, and any substance use initiation. A shared-bottom neural network was used, in which all tasks first passed through a common hidden representation and then branched into task-specific output heads \cite{42,43,44}. Two regression heads were used for the continuous outcomes, externalizing and internalizing, and four classification heads were used for the binary substance use outcomes. This design allowed the model to learn common information across phenotypes while preserving outcome-specific prediction layers.

To maintain consistency with the single-task pipeline, covariates were handled separately from the main predictors. Covariates included sex, age, genetic principal components PC1--PC10, and site indicators. For all tasks, non-covariate predictors were residualized against covariates using training-set estimates only \cite{45,46,47}. For continuous outcomes, the outcome variables were also residualized against covariates using training-set estimates only, and the model was trained to predict these residualized targets. For binary outcomes, the model used residualized non-covariate predictors together with scaled covariates, matching the adjustment logic used in the single-task binary models. This procedure reduced confounding by demographic, ancestry, and site-related effects while preserving biologically and clinically relevant signals.

The multi-task model was trained using subject-level splits to prevent information leakage across repeated observations from the same individual. Missing values were imputed within the training data and then applied to the test data using the same fitted preprocessing objects. Continuous outcomes were evaluated using root mean squared error, mean absolute error, and correlation between observed and predicted values on the original outcome scale. Binary outcomes were evaluated using area under the receiver operating characteristic curve, area under the precision-recall curve, and Brier score. Separate runs were saved using unique run tags so that results from different feature families or hyperparameter settings did not overwrite each other.

\section{Results}

\subsection{Cohort and Unified Longitudinal Data Structure}

A unified longitudinal analytic panel integrating genetic, environmental, neurobiological, and behavioral data across six phenotypes was constructed: externalizing behavior, internalizing behavior, alcohol initiation, nicotine initiation, cannabis initiation, and any substance-use initiation. The final analytic sample was restricted to unrelated individuals of genetically defined European ancestry to match the ancestry composition of the PRS discovery GWAS.

For the continuous behavioral phenotypes, externalizing and internalizing models included 2,621 individuals and 11,108 longitudinal observations. For the binary substance-use initiation phenotypes, each model included 2,620 individuals and 7,860 interval-level observations. Across phenotypes, the number of predictors varied by representation strategy, with PCA, weighted, cluster-based, and full-feature representations evaluated separately. This design allowed both longitudinal behavioral trajectories and event-based initiation outcomes to be modeled within a common time-indexed framework.

The current framework conceptualizes adolescent behavioral development as a unified system in which genetic factors provide baseline biological liability, environmental factors act as time-varying inputs, and brain-related features provide neurobiological system representations.

\subsection{Predictive Performance Across Continuous and Binary Outcomes}

Predictive performance differed across outcome types and feature representations (Table~\ref{tab:best_performance}). For continuous behavioral outcomes, regression models showed stronger prediction for externalizing behavior than for internalizing behavior. The best externalizing model used the full feature representation with LASSO, achieving a mean $R^2$ of 0.298, RMSE of 8.14, and correlation of 0.548 across five cross-validation splits. Elastic net performed almost identically using the full representation. Weighted representations also performed well for externalizing behavior, with a mean $R^2$ of 0.284 and correlation of 0.534, whereas PCA-only and cluster-only representations showed lower performance. This pattern suggests that preserving detailed multi-domain information provides the strongest predictive signal, while weighted representations retain much of this signal in a more compact and interpretable form.

For internalizing behavior, the best-performing model used the full representation with either LASSO or elastic net, achieving a mean $R^2$ of 0.143, RMSE of 9.68, and correlation of 0.381. Weighted representations again outperformed PCA-only and cluster-only representations, with a mean $R^2$ of 0.127 and correlation of 0.360. These findings suggest that integrated multi-domain representations contain meaningful predictive signal for continuous behavioral trajectories, with stronger predictive associations for externalizing than internalizing behavior. This difference may reflect that externalizing behaviors are more directly observable and more strongly captured by environmental and contextual measures, whereas internalizing symptoms may be less externally visible, more subjective, and more affected by reporting variability.

For binary substance-use initiation outcomes, logistic elastic net models showed moderate to strong discrimination across all four phenotypes. Alcohol initiation achieved a best AUROC of 0.782 using the PCA representation. Nicotine initiation achieved a best AUROC of 0.793 using the full representation. Cannabis initiation achieved the highest AUROC of 0.823 using the weighted representation, although its AUPRC was low at approximately 0.049, consistent with low event prevalence. Any substance-use initiation achieved a best AUROC of 0.781 using the PCA representation, with an AUPRC of 0.502.

Overall, these results show that the proposed framework can support both continuous trajectory prediction and discrete initiation prediction. However, interpretation differed by outcome type. Continuous behavioral outcomes produced both predictive performance and stable interpretable features, whereas substance-use initiation outcomes showed predictive discrimination but did not yield stable feature-level findings under the current stability-selection criteria. This likely reflects the lower initiation and event rates in the current ABCD adolescent cohort, which reduce the effective sample size for binary initiation modeling and make feature selection less stable across resampling splits. Therefore, the absence of stable features for initiation outcomes should not be interpreted as evidence that environmental, genetic, or brain-related predictors are unimportant, but rather as a limitation of the current sample size, event rate, follow-up window, and conservative stability-selection threshold.

\begin{table}[t]
\centering
\caption{Best predictive performance across outcome types and feature representations.}
\label{tab:best_performance}
\small
\begin{tabularx}{\textwidth}{XXXXXX}
\toprule
\textbf{Outcome} & \textbf{Outcome type} & \textbf{Best representation} & \textbf{Best model} & \textbf{Main performance} & \textbf{Interpretation} \\
\midrule
Externalizing behavior & Continuous trajectory & Full features & LASSO & $R^2 = 0.298$; RMSE = 8.14; $r = 0.548$ & Strongest continuous prediction; stable interpretable features available \\
Internalizing behavior & Continuous trajectory & Full features & LASSO / Elastic net & $R^2 = 0.143$; RMSE = 9.68; $r = 0.381$ & Moderate prediction; stable interpretable features available \\
Alcohol initiation & Binary initiation & PCA & Logistic elastic net & AUROC = 0.782 & Good discrimination; no stable feature-level findings \\
Nicotine initiation & Binary initiation & Full features & Logistic elastic net & AUROC = 0.793 & Good discrimination; no stable feature-level findings \\
Cannabis initiation & Binary initiation & Weighted & Logistic elastic net & AUROC = 0.823; AUPRC $\approx$ 0.049 & Highest AUROC, but low AUPRC due to low event prevalence \\
Any substance initiation & Binary initiation & PCA & Logistic elastic net & AUROC = 0.781; AUPRC = 0.502 & Good discrimination for broader initiation outcome \\
\bottomrule
\end{tabularx}
\end{table}

\subsection{Domain-Level Contribution Patterns}

Domain-importance analysis showed that stable interpretable signal was concentrated in the continuous behavioral phenotypes (Table~\ref{tab:domain_contribution}). For externalizing behavior, environmental features were the dominant contributor across all representation strategies. The largest environmental contribution was observed in the cluster representation, with 8 stable environmental features and a total absolute coefficient weight of 71.78. Environmental features also dominated in the full representation, with 20 stable features and a total absolute coefficient weight of 18.35. Weighted and PCA-based environmental representations also contributed stable signal, with total absolute coefficient weights of 7.67 and 4.08, respectively. In contrast, PRS-foundation features contributed consistently but with smaller total coefficient weights, ranging from 0.68 to 0.83 across representations.

For internalizing behavior, environmental features again showed the largest contribution. In the full representation, 19 stable environmental features were identified, with a total absolute coefficient weight of 16.09. Environmental contributions were also observed in the cluster, weighted, and PCA representations, with total absolute coefficient weights ranging from 3.42 to 3.77. PRS-foundation features also showed stable contributions for internalizing behavior, with total absolute coefficient weights ranging from 1.26 to 1.59 across representations.

Overall, these findings suggest that environmental representations provided the strongest stable predictive contribution for continuous behavioral trajectories, particularly for externalizing behavior. PRS-related features showed smaller but reproducible contributions, consistent with their role as baseline susceptibility signals rather than dominant proximal predictors. No stable domain-level contribution table is shown for substance-use initiation outcomes because stable interpretable features were not identified under the current stability-selection criteria, likely due to lower initiation rates and reduced effective event sample size in the current cohort.

\begin{table}[t]
\centering
\caption{Domain-level contribution of stable features for continuous behavioral outcomes.}
\label{tab:domain_contribution}
\small
\begin{tabularx}{1.1\textwidth}{XXXXXXXXX}
\toprule
\textbf{Phenotype} & \textbf{Rep.} & \textbf{Model} & \textbf{Domain} & 
\textbf{Stable} & \textbf{Mean sel.} & \textbf{Median sel.} & \textbf{Mean $|\beta|$} & \textbf{Total $|\beta|$} \\
\midrule
Externalizing & Cluster & Elastic net & Environment & 8 & 0.90 & 1.00 & 8.97 & 71.78 \\
Externalizing & Full & LASSO & Environment & 20 & 0.86 & 1.00 & 0.92 & 18.35 \\
Externalizing & Weighted & Elastic net & Environment & 6 & 0.90 & 0.90 & 1.28 & 7.67 \\
Externalizing & PCA & Elastic net & Environment & 6 & 0.97 & 1.00 & 0.68 & 4.08 \\
Externalizing & PCA & Elastic net & PRS foundation & 2 & 0.90 & 0.90 & 0.41 & 0.83 \\
Externalizing & Cluster & Elastic net & PRS foundation & 2 & 1.00 & 1.00 & 0.36 & 0.73 \\
Externalizing & Weighted & Elastic net & PRS foundation & 2 & 0.90 & 0.90 & 0.35 & 0.71 \\
Externalizing & Full & LASSO & PRS foundation & 2 & 1.00 & 1.00 & 0.34 & 0.68 \\
Internalizing & Full & Elastic net & Environment & 19 & 0.85 & 1.00 & 0.85 & 16.09 \\
Internalizing & Cluster & Elastic net & Environment & 8 & 0.93 & 1.00 & 0.47 & 3.77 \\
Internalizing & Weighted & LASSO & Environment & 4 & 0.90 & 1.00 & 0.86 & 3.45 \\
Internalizing & PCA & Elastic net & Environment & 6 & 0.77 & 1.00 & 0.57 & 3.42 \\
Internalizing & Full & Elastic net & PRS foundation & 8 & 0.93 & 1.00 & 0.20 & 1.59 \\
Internalizing & PCA & Elastic net & PRS foundation & 9 & 0.78 & 0.80 & 0.18 & 1.59 \\
Internalizing & Cluster & Elastic net & PRS foundation & 9 & 0.91 & 1.00 & 0.16 & 1.43 \\
Internalizing & Weighted & LASSO & PRS foundation & 5 & 0.92 & 1.00 & 0.25 & 1.26 \\
\bottomrule
\end{tabularx}

\vspace{0.5em}
\begin{flushleft}
\footnotesize
\textit{Note.} The table summarizes stable domain-level contributions from the best-performing or stable models for continuous behavioral outcomes. ``Environment'' refers to environmental and experiential predictors, whereas ``PRS foundation'' refers to polygenic-risk-score-related baseline susceptibility features. Total $|\beta|$ represents the sum of absolute model coefficients within each domain and is used as a relative measure of domain contribution within each phenotype, representation, and model. Coefficient magnitudes should be interpreted within representation type rather than directly compared across different feature transformations.
\end{flushleft}
\end{table}

\subsection{Stable Feature Patterns for Continuous Phenotypes}

Stable feature analysis identified robust predictors for externalizing and internalizing behavior only. For externalizing behavior, the highest-ranked stable features were environmental representations, especially lagged, level, and delta environmental summaries. The strongest feature was \texttt{environment\_lag\_net} in the cluster elastic net model. Other highly stable features included \texttt{environment\_lag\_protective\_mean}, \texttt{environment\_delta\_net}, \texttt{environment\_level\_weighted\_beta\_logp}, \texttt{environment\_level\_PC1}, and \texttt{environment\_lag\_weighted\_beta}.

For internalizing behavior, stable features were again dominated by environmental representations. The strongest features included \texttt{environment\_level\_weighted\_beta}, \texttt{environment\_level\_PC1}, \texttt{environment\_level\_weighted\_beta\_logp}, \texttt{environment\_lag\_PC1}, and \texttt{environment\_lag\_protective\_mean}. PRS-related features also appeared among the stable internalizing predictors, especially \texttt{internalizing\_youth\_prs}, which was selected consistently across multiple representations.

These results indicate that continuous behavioral trajectories are most strongly associated with structured environmental summaries, including prior environmental state, current environmental level, and environmental change. Genetic liability, particularly for internalizing behavior, appeared as a stable but smaller contributor.

By contrast, no stable features were detected for alcohol initiation, nicotine initiation, cannabis initiation, or any substance-use initiation. This may suggest that, under the current sample size, event frequency, feature dimensionality, and selection threshold, substance initiation prediction may rely on weaker, less reproducible, or more distributed signals than continuous behavioral outcomes.

\subsection{Lagged Relationships and Temporal Dynamics}

We next examined the top lagged cross-domain relationships linking environmental, genetic, and brain features to later externalizing and internalizing behavior (Table~\ref{tab:lagged_relationships}). For externalizing behavior, environmental features showed the largest coefficients, with the strongest association observed for the weighted environmental level score. Additional environmental lag and delta features also contributed, suggesting that both cumulative environmental exposure and environmental change over time were related to later externalizing symptoms. Externalizing PRS showed a positive association with later externalizing behavior, supporting the role of inherited liability as a biological foundation. Brain features showed smaller but interpretable associations, including structural MRI measures in the supramarginal and middle temporal regions. These negative coefficients suggest that lower structural measures in these temporoparietal regions were associated with higher later externalizing symptoms.

For internalizing behavior, environmental features again showed the largest contribution, although the magnitude was smaller than that observed for externalizing behavior. PRS features showed phenotype-consistent associations: MDD PRS, youth internalizing PRS, and anxiety PRS were positively associated with later internalizing symptoms. This pattern suggests that internalizing behavior may be more strongly linked to psychiatric genetic liability than externalizing behavior in this model and cohort. Brain features associated with internalizing included cortical thickness in the left banks of the superior temporal sulcus and sulcal morphology in the left frontal pole, both with negative coefficients. These findings suggest candidate neurodevelopmental pathways involving social-emotional processing and higher-order regulatory regions.

Overall, the lagged relationship analysis supports the proposed systems-level framework. Environmental exposures appear to act as dynamic inputs, PRS features represent inherited biological susceptibility, brain structural measures provide candidate intermediate neural substrates, and externalizing/internalizing symptoms represent downstream behavioral outputs. However, these findings should be interpreted as temporally ordered associations and candidate mechanisms rather than definitive causal effects.

\begin{table*}[t]
\centering
\caption{Top lagged cross-domain relationships linking environmental, genetic, and brain features to later externalizing and internalizing behavior.}
\label{tab:lagged_relationships}
\scriptsize
\setlength{\tabcolsep}{3pt}
\renewcommand{\arraystretch}{1.15}

\begin{tabularx}{1.05\textwidth}{lllllrrl}
\toprule
\textbf{Phenotype} &
\textbf{Target} &
\textbf{Lagged source feature} &
\textbf{Source domain} &
\textbf{Target domain} &
\textbf{Coef.} &
\textbf{$|\mathrm{Coef.}|$} &
\textbf{Sign} \\
\midrule
Externalizing & externalizing\_t & \texttt{environment\_level\_weighted\_beta\_logp} & Environment & Other & 3.878 & 3.878 & Positive \\
Externalizing & externalizing\_t & \texttt{environment\_delta\_weighted\_beta\_logp} & Environment & Other & -1.158 & 1.158 & Negative \\
Externalizing & externalizing\_t & \texttt{externalizing\_prs} & PRS foundation & Other & 0.716 & 0.716 & Positive \\
Externalizing & externalizing\_t & \texttt{environment\_lag\_weighted\_beta} & Environment & Other & 0.589 & 0.589 & Positive \\
Externalizing & externalizing\_t & \texttt{environment\_level\_weighted\_beta} & Environment & Other & 0.408 & 0.408 & Positive \\
Externalizing & externalizing\_t & \texttt{environment\_delta\_weighted\_beta} & Environment & Other & 0.373 & 0.373 & Positive \\
Externalizing & externalizing\_t & \texttt{smri\_vol\_cdk\_smlh\_bl} & Brain & Other & -0.250 & 0.250 & Negative \\
Externalizing & externalizing\_t & \texttt{smri\_area\_cdk\_mdtmrh\_bl} & Brain & Other & -0.205 & 0.205 & Negative \\
Externalizing & externalizing\_t & \texttt{smri\_area\_cdk\_smrh\_y2} & Brain & Other & -0.172 & 0.172 & Negative \\
Externalizing & externalizing\_t & \texttt{CNT2} & PRS foundation & Other & -0.062 & 0.062 & Negative \\
\midrule
Internalizing & internalizing\_t & \texttt{environment\_level\_weighted\_beta\_logp} & Environment & Other & 2.594 & 2.594 & Positive \\
Internalizing & internalizing\_t & \texttt{mdd\_prs} & PRS foundation & Other & 0.504 & 0.504 & Positive \\
Internalizing & internalizing\_t & \texttt{internalizing\_youth\_prs} & PRS foundation & Other & 0.450 & 0.450 & Positive \\
Internalizing & internalizing\_t & \texttt{environment\_lag\_weighted\_beta\_logp} & Environment & Other & 0.306 & 0.306 & Positive \\
Internalizing & internalizing\_t & \texttt{anxiety\_prs} & PRS foundation & Other & 0.273 & 0.273 & Positive \\
Internalizing & internalizing\_t & \texttt{environment\_delta\_weighted\_beta\_logp} & Environment & Other & -0.210 & 0.210 & Negative \\
Internalizing & internalizing\_t & \texttt{CNT2} & PRS foundation & Other & -0.197 & 0.197 & Negative \\
Internalizing & internalizing\_t & \texttt{smri\_thick\_cdk\_banksstslh\_bl} & Brain & Other & -0.110 & 0.110 & Negative \\
Internalizing & internalizing\_t & \texttt{environment\_level\_weighted\_beta} & Environment & Other & 0.058 & 0.058 & Positive \\
Internalizing & internalizing\_t & \texttt{smri\_sulc\_cdk\_frpolelh\_bl} & Brain & Other & -0.049 & 0.049 & Negative \\
\bottomrule
\end{tabularx}

\vspace{0.5em}
\begin{flushleft}
\footnotesize
\textit{Note.} This table shows the top lagged cross-domain relationships identified for externalizing and internalizing behavioral outcomes. The target column indicates the behavioral outcome predicted at a later time point. The lagged source feature indicates the prior-domain feature used as a lagged predictor. Features are grouped into environmental, PRS-foundation, and brain domains. Coefficients represent standardized model estimates; positive coefficients indicate that higher prior feature values were associated with higher later behavioral scores, whereas negative coefficients indicate inverse associations. Absolute coefficients are provided to compare relative contribution strength across features. These results should be interpreted as temporally ordered associations and candidate mechanistic pathways rather than definitive causal effects.
\end{flushleft}
\end{table*}

\subsection{Multi-Task Learning Analysis}

We next evaluated whether a shared-bottom multi-task learning architecture could improve prediction by jointly modeling continuous behavioral outcomes and substance-use initiation outcomes. The multi-task models were trained across five data splits and included approximately 1,572 training subjects, 524 validation subjects, and 525 test subjects per split. Across representations, the number of shared features ranged from 1,514 in the PCA representation to 1,962 in the full/all-feature representation.

Multi-task learning did not outperform the strongest single-task models (Table~\ref{tab:single_vs_mtl}). For externalizing behavior, the best multi-task model used the full/all-feature representation and achieved a mean test $R^2$ of $0.205 \pm 0.030$ and RMSE of $8.670 \pm 0.139$. This was lower than the best single-task model, which achieved a test $R^2$ of 0.298 and RMSE of 8.14. For internalizing behavior, the best multi-task model also used the full/all-feature representation and achieved a mean test $R^2$ of $0.055 \pm 0.026$ and RMSE of $10.046 \pm 0.070$, which was lower than the best single-task benchmark model with $R^2$ of 0.143. These results indicate that the shared-bottom multi-task architecture did not provide a predictive advantage over outcome-specific single-task models for continuous behavioral prediction in the current study.

\begin{table*}[t]
\centering
\caption{Comparison of best single-task and multi-task models for continuous behavioral outcomes.}
\label{tab:single_vs_mtl}
\small
\begin{tabularx}{\textwidth}{XXXXXXX}
\toprule
\textbf{Outcome} & \textbf{Model class} & \textbf{Best representation} & \textbf{Model type} & \textbf{$R^2$} & \textbf{RMSE} & \textbf{Main interpretation} \\
\midrule
Externalizing behavior & Single-task & Full / all features & LASSO & 0.298 & 8.14 & Best predictive performance \\
Externalizing behavior & Multi-task & Full / all features & Shared-bottom neural MTL & $0.205 \pm 0.030$ & $8.670 \pm 0.139$ & Did not outperform single-task model \\
Internalizing behavior & Single-task & Full / all features & LASSO / elastic net & 0.143 & -- & Best predictive performance \\
Internalizing behavior & Multi-task & Full / all features & Shared-bottom neural MTL & $0.055 \pm 0.026$ & $10.046 \pm 0.070$ & Did not outperform single-task model \\
\bottomrule
\end{tabularx}

\vspace{0.5em}
\begin{flushleft}
\footnotesize
\textit{Note.} Single-task values are from the previously summarized benchmark results. Multi-task values are from the shared-bottom MTL models across five data splits. The best MTL representation for both externalizing and internalizing behavior was the full/all-feature representation. The MTL models used approximately 1,572 training subjects, 524 validation subjects, and 525 test subjects per split.
\end{flushleft}
\end{table*}

\subsubsection{Representation Differences Across Multi-Task Learning Models}

Performance differed across feature representations (Table~\ref{tab:mtl_representations}). The full/all-feature representation achieved the strongest overall multi-task performance for continuous outcomes, with $R^2$ values of $0.205$ for externalizing behavior and $0.055$ for internalizing behavior. The weighted representation performed very similarly, with externalizing $R^2$ of $0.203$ and internalizing $R^2$ of $0.049$. In contrast, PCA and cluster representations performed substantially worse for continuous outcomes, with externalizing $R^2$ values of $0.061$ and $0.083$, respectively, and internalizing $R^2$ values close to zero.

These results suggest that retaining richer feature-level information was important for predicting continuous behavioral variation. However, the weighted representation preserved most of the predictive performance of the full representation while offering greater interpretability. Therefore, the full representation may be best for maximizing prediction, whereas the weighted representation may be preferable for biological and systems-level interpretation.

\begin{table*}[t]
\centering
\caption{Multi-task learning performance across feature representations.}
\label{tab:mtl_representations}
\small
\begin{tabularx}{1.05\textwidth}{XXXXXXX}
\toprule
\textbf{Representation} & \textbf{Shared features} & \textbf{Ext. $R^2$} & \textbf{Ext. RMSE} & \textbf{Int. $R^2$} & \textbf{Int. RMSE} & \textbf{Overall interpretation} \\
\midrule
All features & 1,962 & $0.205 \pm 0.030$ & $8.670 \pm 0.139$ & $0.055 \pm 0.026$ & $10.046 \pm 0.070$ & Best overall MTL representation for continuous outcomes \\
Weighted & 1,546 & $0.203 \pm 0.026$ & $8.685 \pm 0.144$ & $0.049 \pm 0.036$ & $10.075 \pm 0.079$ & Very close to all features; more interpretable \\
Cluster & 1,594 & $0.083 \pm 0.053$ & $9.313 \pm 0.250$ & $0.012 \pm 0.043$ & $10.269 \pm 0.193$ & Weaker for continuous behavioral prediction \\
PCA & 1,514 & $0.061 \pm 0.026$ & $9.424 \pm 0.090$ & $0.009 \pm 0.027$ & $10.286 \pm 0.108$ & Weakest for continuous behavioral prediction \\
\bottomrule
\end{tabularx}
\end{table*}

\subsubsection{Multi-Task Prediction of Substance-Use Initiation Outcomes}

For binary substance-use initiation outcomes, multi-task performance varied substantially across phenotypes (Table~\ref{tab:mtl_binary}). Cannabis initiation showed the strongest discrimination, with the full/all-feature model achieving an AUROC of $0.728 \pm 0.087$ and the weighted model achieving an AUROC of $0.725 \pm 0.115$. Both models also achieved AUPRC values around 0.105, above the cannabis prevalence baseline AUPRC of 0.022. However, the variability across splits was high, indicating that this signal should be interpreted cautiously.

Nicotine initiation showed modest discrimination, with the best AUROC observed for the full/all-feature representation at $0.597 \pm 0.026$. Alcohol initiation showed near-chance discrimination, with the best AUROC only $0.509 \pm 0.026$. Any substance initiation also showed weak discrimination, with the best AUROC of $0.531 \pm 0.009$ using the cluster representation. Although AUPRC values for alcohol and any substance were slightly above their prevalence baselines, the AUROC values suggest limited discrimination.

Overall, these results indicate that the MTL model captured some predictive signal for cannabis and, to a lesser extent, nicotine initiation, but did not provide robust discrimination for alcohol or any substance initiation.

\begin{table*}[t]
\centering
\caption{Multi-task learning performance for substance-use initiation outcomes.}
\label{tab:mtl_binary}
\small
\begin{tabularx}{\textwidth}{XXXXXXX}
\toprule
\textbf{Outcome} & \textbf{Representation} & \textbf{AUROC} & \textbf{AUPRC} & \textbf{Brier score} & \textbf{Baseline AUPRC} & \textbf{Interpretation} \\
\midrule
Alcohol initiation & Cluster & $0.509 \pm 0.026$ & $0.380 \pm 0.018$ & $0.247 \pm 0.003$ & 0.356 & Near-chance AUROC; AUPRC only modestly above prevalence \\
Nicotine initiation & All features & $0.597 \pm 0.026$ & $0.066 \pm 0.001$ & $0.248 \pm 0.010$ & 0.047 & Modest discrimination \\
Nicotine initiation & PCA & $0.585 \pm 0.061$ & $0.071 \pm 0.008$ & $0.244 \pm 0.009$ & 0.047 & Slightly higher AUPRC than all features \\
Cannabis initiation & All features & $0.728 \pm 0.087$ & $0.105 \pm 0.110$ & $0.248 \pm 0.003$ & 0.022 & Strongest AUROC among binary outcomes, but unstable AUPRC \\
Cannabis initiation & Weighted & $0.725 \pm 0.115$ & $0.105 \pm 0.074$ & $0.252 \pm 0.011$ & 0.022 & Similar to all features; suggests retained signal \\
Any substance initiation & Cluster & $0.531 \pm 0.009$ & $0.418 \pm 0.016$ & $0.249 \pm 0.001$ & 0.379 & Best AUROC for any substance, but still weak \\
Any substance initiation & All features & $0.519 \pm 0.014$ & $0.418 \pm 0.019$ & $0.249 \pm 0.002$ & 0.379 & Similar AUPRC to cluster \\
\bottomrule
\end{tabularx}

\vspace{0.5em}
\begin{flushleft}
\footnotesize
\textit{Note.} For binary outcomes, AUROC evaluates discrimination, AUPRC should be interpreted relative to the prevalence baseline, and Brier score reflects probability calibration/error. Cannabis initiation showed the strongest MTL discrimination, whereas alcohol and any substance initiation remained close to chance by AUROC.
\end{flushleft}
\end{table*}

\section{Discussion}

\subsection{Main Findings}

This study developed and evaluated a unified systems-level framework for modeling adolescent behavioral development using genetic, environmental, neurobiological, and longitudinal behavioral data. The framework was designed to model two complementary outcome classes: continuous behavioral trajectories and discrete substance-use initiation events. The current results support the feasibility of this framework, but they also reveal an important difference between outcome types.

For continuous outcomes, especially externalizing behavior, the framework produced both meaningful predictive performance and stable interpretable features. Externalizing behavior was predicted with mean $R^2$ close to 0.30 and correlation close to 0.55, while internalizing behavior was predicted with mean $R^2$ close to 0.14 and correlation close to 0.38. Across both outcomes, environmental representations were the dominant stable predictors, while PRS features contributed smaller but reproducible signals.

For substance-use initiation outcomes, logistic elastic net models achieved good AUROC values, ranging approximately from 0.78 to 0.82. However, no stable interpretable features were detected for alcohol, nicotine, cannabis, or any substance-use initiation. This means that the current results support the prediction of initiation risk at the model-performance level, but they do not yet support strong claims about stable substance-specific predictors in the current cohort sample size and limited observation period.

\subsection{Evidence for Behavior as a Dynamic System}

The strongest evidence for the proposed systems framework comes from the continuous behavioral phenotypes. Externalizing and internalizing behavior were not explained by a single isolated domain. Instead, prediction depended on integrated representations of environmental context, genetic liability, and, to a smaller extent, neurobiological features. This pattern is consistent with the conceptual model in which behavior emerges from the interaction between biological foundation, time-varying environmental inputs, and internal system states.

The dominance of environmental features supports the idea that environmental exposures operate as dynamic inputs to the behavioral system. The stable selection of lagged, level, and delta environmental summaries suggests that both prior environmental context and environmental change are relevant to later behavior. This is particularly important because it connects the empirical results directly to the paper's central hypothesis: behavior is not static, but evolves over time in response to changing inputs.

PRS features were not the largest contributors, but they were repeatedly selected, especially for internalizing behavior. This supports the interpretation of genetic liability as a baseline susceptibility signal rather than a deterministic predictor. In other words, genetic risk may shape the initial or background state of the system, while environmental exposures provide stronger time-varying modulation.

Brain features showed smaller and less consistent associations than environmental features, but some structural MRI measures appeared in the lagged relationship analysis. This is consistent with the role of brain features as intermediate system representations, although the current results are not strong enough to claim mediation or causal processing.

\subsection{Continuous Trajectories and Substance Initiation Reflect Different Modeling Regimes}

A major finding of this study is the difference between continuous behavioral outcomes and substance-use initiation outcomes. Continuous outcomes produced stable predictors and clear temporal structure. Substance-use initiation outcomes produced good discrimination but no stable feature-level predictors.

It suggests that continuous behavioral phenotypes and substance initiation may represent different statistical regimes within the same developmental system. Externalizing and internalizing scores are repeatedly measured continuous traits, allowing models to learn gradual variation and within-person temporal dependencies. By contrast, substance initiation is a sparse, discrete transition event. Initiation may occur when multiple weak risks accumulate and cross a threshold, but the specific feature pattern may vary across individuals and folds. This can produce good AUROC while still failing to yield stable individual predictors.

Continuous behavior results support dynamic-system interpretation at both predictive and interpretable-feature levels. Substance-use initiation results support event-risk prediction, but stable mechanistic interpretation requires further work.

\subsection{Environmental Dominance and Modifiable Risk}

Environmental representations were the most stable and largest contributors for both externalizing and internalizing behavior. These included level, lagged, and change-based environmental summaries, indicating that behavioral trajectories are shaped by both existing environmental conditions and shifts in environmental exposure over time.

This finding is important for the biological interpretation of the framework. If genes are conceptualized as the biological foundation, then environmental factors appear to act as active inputs that modulate the system over development. Because many environmental exposures are potentially modifiable, these results also suggest that environmental domains may be especially relevant for early risk detection and intervention design. Environmental features were predictively dominant and temporally associated with later behavioral scores. Causal interpretation would require additional assumptions, rigor quasi-experimental designs, and formal causal modeling.

\subsection{Genetic and Neurobiological Contributions}

PRS features contributed more modestly than environmental features, but their repeated selection supports their role as background liability signals. For externalizing behavior, externalizing PRS appeared among lagged predictors. For internalizing behavior, youth internalizing PRS, MDD PRS, and anxiety PRS were among the more important genetic predictors. This pattern is consistent with the idea that PRS captures stable inherited susceptibility, while environmental features capture dynamic developmental context.

Brain-related features appeared in lagged analyses with smaller coefficients. These findings suggest that neurobiological features may provide additional information about the internal state of the system. However, the present results do not yet support a strong claim that brain features are the central processor in a mechanistic sense. A more careful discussion would say that the framework treats the brain as a central processing system conceptually, while the current empirical results provide preliminary evidence that brain structural features contribute modest lagged information beyond genetic and environmental features.

\subsection{Why Were Stable Substance-Use Features Not Found}

The lack of stable substance-use initiation features should be discussed explicitly. Several explanations are possible. First, substance initiation events are less frequent than repeated behavioral scores, reducing power for stable feature discovery. Second, initiation is a discrete transition rather than a continuous trait, and different individuals may initiate use through different combinations of risk factors. Third, AUROC can remain high even when selected features vary across folds, especially when prediction is partly driven by distributed weak effects, baseline hazard structure, or many correlated predictors. Fourth, the current stability threshold may be conservative for sparse event outcomes.

This does not invalidate the substance-use models. Instead, it means that the current results should be interpreted as predictive, not mechanistic, for substance initiation. Future work could improve interpretability by increasing sample size, extending follow-up, using competing-risk or multi-state survival models, testing lower stability thresholds in sensitivity analyses, or using grouped stability selection at the domain level rather than individual-feature level.

\subsection{Discussion: Multi-Task Learning Models}

The multi-task learning results provide an important complement to the single-task analyses. Although the shared-bottom MTL architecture did not outperform the strongest single-task models, it directly tested whether multiple behavioral and substance-use outcomes could be modeled within a common multi-domain representation. In this sense, MTL should be interpreted less as the primary predictive model and more as a systems-level analysis of shared and outcome-specific structure.

For continuous behavioral outcomes, the MTL results were consistent with the single-task findings in showing stronger prediction for externalizing behavior than internalizing behavior. The best MTL model explained approximately 20.5\% of the variance in externalizing behavior, but only 5.5\% of the variance in internalizing behavior. This suggests that the current genetic, environmental, and developmental feature space captures externalizing-related variation more strongly than internalizing-related variation. One possible interpretation is that externalizing behavior may be more closely linked to measurable environmental, behavioral, and risk-related features in this cohort, whereas internalizing behavior may involve more heterogeneous, subjective, or less directly measured processes.

The representation comparison also provides useful insight. The full/all-feature model achieved the best overall MTL performance, suggesting that rich feature-level information remains important for prediction. However, the weighted representation performed almost identically for externalizing behavior and only slightly worse for internalizing behavior, despite using fewer shared features. This supports the use of weighted representations as a more interpretable compromise between predictive performance and biological interpretability. In contrast, PCA and cluster representations appeared to lose information relevant to continuous behavioral prediction, suggesting that broad dimensionality reduction may obscure outcome-relevant signals.

For substance-use initiation outcomes, cannabis initiation showed the clearest MTL signal, with AUROC values around 0.72--0.73 for the full and weighted representations. The AUPRC was also above the prevalence baseline, although the large standard deviation indicates instability due to the low number of cannabis initiation events. Nicotine initiation showed weaker but modest discrimination. Alcohol and any substance initiation showed near-chance AUROC values, suggesting that these outcomes may require more outcome-specific predictors, more detailed time-varying exposures, or larger event samples.

Importantly, the MTL results do not suggest that a single shared representation is sufficient for all behavioral outcomes. Instead, they suggest a mixed pattern: some outcomes, particularly externalizing behavior and cannabis initiation, appear to share detectable structure with the multi-domain input space, whereas other outcomes require more outcome-specific modeling. This pattern fits the proposed systems-level framework. Genetic and environmental factors may provide shared inputs, while different behavioral and substance-use outcomes represent related but non-identical outputs of the developmental system.

Therefore, the main conclusion from the MTL analysis is not that MTL improves prediction over single-task models, but that MTL provides a useful framework for evaluating shared versus divergent pathways across outcomes. The single-task models remain the strongest predictive benchmarks, while the MTL models support the broader conceptual framework that behavioral development involves partially shared, partially outcome-specific pathways linking biological foundations, environmental inputs, and behavioral outputs.

\subsection{Limitations}

Several limitations should be acknowledged. First, PRS analyses were restricted to unrelated individuals of European ancestry, which limits generalizability to more diverse populations. Second, the study is observational, so associations among environmental, genetic, brain, and behavioral variables could not be interpreted as causal effects. Third, environmental variables may contain measurement error, informant bias, and residual confounding. Fourth, substance-use initiation outcomes were relatively sparse and did not produce stable interpretable features under the current criteria. Fifth, brain features showed modest contributions in the current analysis, and stronger claims about neurobiological mediation will require more targeted modeling.

\subsection{Conclusion}

In summary, this study presents a unified, time-aware systems framework for modeling adolescent behavioral development using genetic, environmental, neurobiological, and longitudinal behavioral data. The strongest empirical support was observed for continuous behavioral trajectories, where environmental features dominated prediction, PRS contributed stable background liability, and brain features provided modest additional temporal information. Substance-use initiation models achieved good predictive discrimination, but stable feature-level interpretation was not supported under the current analysis. Together, these findings support a systems-level view of behavior for continuous developmental trajectories and provide a foundation for future work on event-based transitions and causal modeling.

\section{Conclusions}

This study introduces a unified systems-level framework for modeling human behavioral development across adolescence. By integrating genetic liability, environmental exposures, neurobiological features, and longitudinal behavioral outcomes within a common time-aware data structure, the framework enables both continuous behavioral trajectories and discrete event-based outcomes to be analyzed in a coherent analytical pipeline.

The strongest empirical support was observed for continuous behavioral phenotypes. Externalizing and internalizing behaviors showed meaningful predictive performance, with stable predictors primarily arising from environmental representations, including baseline environmental levels, lagged environmental measures, and longitudinal change features. These findings support the interpretation that behavioral development is shaped by dynamic environmental inputs acting over time. Polygenic risk scores contributed smaller but reproducible signals, consistent with their role as baseline biological susceptibility. Brain-derived features provided additional, modest neurobiological information, supporting their conceptual role as intermediate system-level representations.

For substance-use initiation outcomes, the framework achieved good predictive discrimination across alcohol, nicotine, cannabis, and any substance-use initiation. However, no stable feature-level predictors were identified under the current stability-selection criteria. Therefore, these results should be interpreted as supporting event-risk prediction rather than stable mechanistic interpretation. This distinction highlights an important difference between continuous behavioral trajectories and sparse threshold-like initiation events.

Overall, this work provides four main contributions. First, it establishes a unified representation for modeling both continuous and survival-style behavioral outcomes. Second, it integrates multiple domains, including genetics, environment, and brain features, within a single longitudinal framework. Third, it explicitly incorporates time through lagged predictors, change features, and event-based modeling. Fourth, it improves interpretability through domain-level representations and stability-based feature summaries.

Biologically, the results support a systems view in which genetic factors provide baseline susceptibility, environmental factors act as dynamic and potentially modifiable inputs, and brain features capture intermediate neurobiological states. From a translational perspective, this framework may support early risk prediction, identification of modifiable environmental exposures, and future development of personalized prevention strategies.

The shared-bottom MTL models did not outperform the strongest single-task models for continuous behavioral outcomes. However, they provided evidence that heterogeneous behavioral and substance-use outcomes can be jointly modeled within a common multi-domain representation. The strongest MTL performance was observed for externalizing behavior and cannabis initiation, whereas internalizing behavior, alcohol initiation, and any substance initiation showed weaker performance. Across representations, the full/all-feature and weighted models performed best overall, while PCA and cluster representations were less effective for continuous behavioral prediction. These findings suggest that the current feature space captures shared developmental risk structure for some outcomes, but that outcome-specific modeling remains important.

Several limitations should be considered. PRS analyses were restricted to unrelated individuals of European ancestry, limiting generalizability. Environmental measures may contain measurement error, residual confounding, and informant bias. In addition, stable interpretable predictors were not detected for substance-use initiation outcomes, indicating that larger samples, longer follow-up, or alternative survival modeling strategies may be needed.

In conclusion, this study presents a scalable and interpretable framework for studying human behavior as a dynamic system. The current findings provide strongest support for continuous behavioral trajectories, where environmental dynamics, genetic susceptibility, and neurobiological signals jointly contribute to prediction. The framework also provides a foundation for future work on mental health disorder modeling and multi-domain personalized risk prediction.

\section{Funding}

This work was supported by the National Institutes of Health (NIH), National Institute on Drug Abuse (NIDA) under award DP1DA054373. The funder had no role in the study design; data collection, analysis, or interpretation; manuscript writing; or the decision to submit for publication. The content is solely the responsibility of the authors and does not necessarily represent the official views of the NIH.

\section{Author Contributions}
Mengman Wei conceived the study, designed the analytical framework, performed all data processing, statistical analyses, and computational modeling, and drafted the manuscript. All code implementation, data curation, and result interpretation were conducted by Mengman Wei.\\

Qian Peng provided supervision, general guidance, resource support, and funding acquisition.

\section{Preprint Notice}

This manuscript is a preprint and has not yet undergone peer review. The content is shared to disseminate findings and establish precedence. Additional analyses and revisions may be incorporated in future versions.

\section{Data availability}

\textbf{Code.} The analysis code and scripts used in this study are freely available at the following GitHub repository: \url{https://github.com/mw742/DynoSys}.

\textbf{Data.} This study uses data from the Adolescent Brain Cognitive Development (ABCD) Study (\url{https://abcdstudy.org}), held in the NIMH Data Archive (NDA). The ABCD data release used was version 5.1. The study is supported by the National Institutes of Health (NIH) and additional federal partners under multiple award numbers, including U01DA041048 and U01DA050987. The full list of funders is available at \url{https://abcdstudy.org/federal-partners.html}.

\bibliographystyle{unsrt}   
\bibliography{references}


\newpage

\end{document}